\documentclass[nofootinbib,superscriptaddress,twocolumn,aps,showpacs, floatfix]{revtex4-1}
\usepackage{epsfig}
\usepackage{graphicx}
\usepackage{dcolumn}
\usepackage{babel}
\usepackage{bm}
\usepackage{amssymb}
\usepackage{amsthm,amsmath}
\usepackage{subfigure}
\usepackage{float}
\usepackage[utf8]{inputenc}
\usepackage{booktabs}
\setcitestyle{authoryear,round}
\usepackage{comment}
\usepackage{hyperref}
\hypersetup{colorlinks=true, citecolor=blue, urlcolor=blue, linkcolor=blue}
\usepackage[dvipsnames, table, xcdraw]{xcolor}
\usepackage[referable]{threeparttablex}
\usepackage{soul}
\usepackage{longtable}
\usepackage{adjustbox}

\usepackage{placeins}
\usepackage{tabularx}

\makeatletter
\renewcommand{\p@subsection}{}
\renewcommand{\p@subsubsection}{}
\makeatother

\begin{document}

%\title{A machine learning model to identify corruption in public procurement contracts: M\'exico a case study.}

\title{A machine learning model to identify corruption in M\'exico's public procurement contracts}

%\title{Identifying corruption in M\'exico´s public procurement: A machine learning approximation}

%\title{Machine learning for predicting corruption in public procurement in M\'exico.}

\author{Andrés Aldana}
\email{andres.aldana@c3.unam.mx}
\affiliation{Centro de Ciencias de la Complejidad, Universidad Nacional Aut\'onoma de M\'exico, Ciudad de M\'exico 04510, M\'exico}

\author{Andrea Falc\'on-Cort\'es}
\email{falcon@icf.unam.mx}
\affiliation{Instituto de Ciencias F\'isicas, Universidad Nacional Aut\'onoma de M\'exico, Morelos 62210, M\'exico}

\author{Hernán Larralde}
\email{hernan@icf.unam.mx}
\affiliation{Instituto de Ciencias F\'isicas, Universidad Nacional Aut\'onoma de M\'exico, Morelos 62210, M\'exico}

\date{\today}

\begin{abstract}
    The costs and impacts of government corruption range from impairing a country's economic growth to affecting its citizens' well-being and safety. Public contracting between government dependencies and private sector instances, referred to as public procurement, is a fertile land of opportunity for corrupt practices, generating substantial monetary losses worldwide. Thus, identifying and deterring corrupt activities between the government and the private sector is paramount. However, due to several factors, corruption in public procurement is challenging to identify and track, leading to corrupt practices going unnoticed. This paper proposes a machine learning model based on an ensemble of random forest classifiers, which we call hyper-forest, to identify and predict corrupt contracts in M\'exico's public procurement data. This method's results correctly detect most of the corrupt and non-corrupt contracts evaluated in the dataset. Furthermore, we found that the most critical predictors considered in the model are those related to the relationship between buyers and suppliers rather than those related to features of individual contracts. Also, the method proposed here is general enough to be trained with data from other countries. Overall, our work presents a tool that can help in the decision-making process to identify, predict and analyze corruption in public procurement contracts.\\
    
{\it Keywords:} Public procurement, corruption, machine learning, hyper-forest, M\'exico.
\end{abstract}

\maketitle

\section{Introduction}

Corruption, defined as the abuse of public power and/or resources for private gains, is widespread across societies, affecting sectors such as healthcare, politics, justice and public procurement \citep{klitgaard1988controlling, rose1996altruism, shleifer1993corruption, mackey2016disease}. The economic cost of corruption is huge; only in 2016 the World Bank estimated that more than \$2.6 trillion USD was lost due to corrupt activities \citep{lima2020predicting}. Even when this number is in its self scandalous, the costs and impacts of corruption go far beyond the monetary issues; corruption distorts public trust, interferes with effective political decision-making, and impairs economic growth \citep{grimes2013contingencies}. 

%For all these severe consequences, corruption is one of the biggest challenges a government has to deal with. \\

In the last few years, the digitalization and access to public data has been gaining ground. Many states have taken advantage of this new \lq\lq technology era" to record their administrative activities, and many have adopted a policy of transparency regarding these data. This increase in access to information may be used to study corruption from new perspectives \citep{radermacher2018official}.

Public procurement is one of the government activities in which usually large amounts of money are invested to contract goods or services from the private sector. Much of this money is lost due to corrupt practices.
%In M\'exico alone, spending in public procurement reached 10\% of the approved federal budget between 2013 and 2020, and almost 2\% of this money was lost in corrupt practices  \citep{falcon2022practices}. 
This draining in public resources through corruption in public procurement occurs around the globe; the World Bank and the Organization for Economic Co-operation and Development consider that the interaction activities between public and private sectors everywhere are always at risk being or becoming corrupt \citep{world2020enhancing,oecd2016preventing}. Besides the consequences enlisted above, corruption in public procurement has an impact on several social levels. It affects the market competition, changes the market dynamics, provokes a misallocation and unequal distribution of resources, and creates barriers to public services and goods. These barriers have an incalculable impact on the citizen's well-being \citep{kohler2018global, mackey2012combating, vian2008review, north2009violence, aidt2016rent, wachs2019social, fazekas2020corruption,meyer2019poder}. However, due to the diversity of possible corrupt operations, the complexity of the contracts involved, the intricate interactions between participants, and the participation/complicity of public officials, corruption in public procurement is challenging to identify, track, predict, and prevent \citep{baldi2016bid, beth2007integrity, oecd2016government, oecd2016preventing}. 

% The most common approach to try to characterize and prevent corruption in public procurement is building risk factors from contract data. The Center for Research on Corruption in Budapest has curated a list of the most relevant indicators of corruption risk in public procurement, and they have studied the efficiency of these risk factors \citep{wachs2019social,fazekas2020corruption, david2020grand,fazekas2013anatomy,fazekas2013corruption,fazekas2016objective,wachs2021corruption}. 
% A usual technique analysis is to use linear, or multi-linear, 
% regression models to find the best predictors of corrupt practices 
% %\citep{charron2017careers,falcon2022practices}. 
% in \cite{falcon2022practices} however, we test four risk factors related to M\'exico's public procurement, finding that isolated risk factors (or even linear combinations of them) result in very poor predictors of corruption. Nevertheless, since the phenomenon of corruption is highly complex, perhaps a better way to predict it would be through non-linear functions of the risk factors.

In this work, we implement a machine learning model based on an ensemble of random forest, which we call {\it hyper-forest}, to detect corrupt contracts in M\'exico's public procurement. For this study, we use the same data analyzed in \cite{falcon2022practices}. In this dataset, besides considering the standard variables of a contract (such as spending, type of contract, etc.), we also include variables such as the buyers' features, descriptors of the buyer/supplier relation, and some of the risk factors proposed in \cite{fazekas2016objective}; all built from the available public data. Our optimal classifier can detect corrupt contracts with an accuracy of 88\% and non-corrupt contracts with 94\% accuracy. We calculate the feature importance to find which variables are the most important for the model and to assess the optimal relative order of the predictors. Then we implement recursive feature elimination to find the set that gives the best classification performance. We find that those variables related to the relationship between buyer/supplier and risk factors are more efficient predictors than those that only describe contract features.

The results in this paper are based solely on the analysis of the data, trying to avoid any political bias, and without use of any prior knowledge about contracts' participants.

\section{Background}

\subsection{Corruption in Public Procurement in Mexico}
%definiciones de corrupcion en contrataciones publicas en sus distintas etapas y el impacto de este tipo corrupcion
Public procurement is the process by which governmental entities spend public money to purchase products and services, having a fundamental role in the economy of any country \citep{oecd2016government}. Ideally, public contracts must guarantee goods and services of the best quality at a minor cost, prioritizing the citizen's benefit over particular interests. Unfortunately, among all the government activities, public procurement is one of the most vulnerable to corruption \citep{oecd2016preventing}; this is due to the large amount of works and services that each government office needs. This translates into thousands of transactions between the government and private companies a
year, where, within the adjudication process of each public purchase, different private interests come together \citep{IMCOmapeando}. Indeed, public procurement becomes corrupt when the election of the products and suppliers is taken following private interests instead of public well-being \citep{worldbankfraud}. 

Public procurement processes comprise all the institutional efforts necessary for the government to acquire a good or service. Generally, these are stipulated in a legal system that regulates the process to be followed step by step \citep{soreide2002corruption}. In M\'exico, this regulation is stipulated in the Law of Public Works and Acquisition of Federal Services \citep{leyfederalservicios}. According to Transparency International, the processes of public procurement are made in five main stages \citep{ticorruptionpp}:
\begin{enumerate}
    \item [1.] {\it Identification of the need}: Recognize the good or service the government agency wants to acquire.
    \item [2.] {\it Method selection and documentation preparation}: Procedure by which the purchase of the previously identified good or service is made (open or restricted contest, or single-bidder). In this stage, the government agency also prepares the documentation to specify the features and requirements of the suppliers.    
    \item [3.] {\it Proposals evaluation and selection}: All the received proposals are evaluated following the features and requirements previously specified, deciding which supplier wins the contract. 
    \item [4.] {\it Contract implementation}: All the activities related to management, monitoring and contract modifications. 
    \item [5.] {\it Final accounting and auditing}: Once the purchasing processes have been completed, an analysis of the suppliers, the contracting dependency and the result of the contracted object are carried out. 
\end{enumerate}

In each of its stages, public procurement is susceptible to corruption. For example, \citet{rose1975economics} discussed the cases when the government does not have a clear preference for the product it wants to buy (stage 1) or when there is one single seller who can provide the good (stage 3). In these cases, incentives exist for suppliers to bribe government officials to either disqualify or ignore their competitors, or to allow them to make extraordinary profits. On the other hand, \citet{soreide2002corruption} analyzes different situations under which corruption may appear at different stages of the process, for example, limiting the call for bids or making the bids' features specific for only one tender (stage 2). In M\'exico, corrupt practices in several of the stages of public procurement have been detected over the years. For example; suppliers that win contracts by presenting fake documentation; companies that do not deliver the goods or services accorded in the contract; shell companies that provide fake receipts to evade taxes or hide embezzlement; suppliers that overcharge their services, breaching their contract, among others. In response, M\'exicos' government has documented lists of the companies that have been caught incurring in any act of corruption. These lists are managed by the Mexican Agency Tax and Mexican Federal Agency for Budget and Public Debt, they are public and we use them in this work to identify the contracts that are suspect of being corrupt. The specifications of these lists are described in Section \ref{sec:methods}.

Corruption in public procurement differs from corruption in other government functions due to the participation of different actors, the specific regulation and the wide margin of discretion in the decision-making. It often involves complex, high-value transactions that offer lucrative opportunities. But even for smaller contracts, this type of spending is often highly vulnerable to corruption, giving rise to the loss of enormous amounts of public resources \citep{mexicanos}. In M\'exico alone, spending in public procurement reached 10\% of the approved federal budget between 2013 and 2020, and almost 2\% of this money was lost in corrupt practices  \citep{falcon2022practices}. The consequences of corruption in public procurement are vast and very damaging, ranging from destabilizing the country's economy to affecting the well-being of its citizens. Thus, it is essential to implement techniques that aid in the quantification and prediction of corrupt practices in any stage of the public procurement process.

\subsection{Approaches to quantify and predict corruption in public procurement}
%importancia de cuantificar esta corrupcion y distintos metodos para intentar cuantificarla
%métodos tanto estadisticos como otros para predecir la corrupción en contrataciones publicas y porque son insuficientes
A common approach to quantify and predict corruption in public procurement is to study the causes of corruption through regression-based analysis considering a wide range of variables. For example, constucting simple linear regression models that correlate the corruption perception index \citep{itcpi} with the features of the country's electoral system \citep{chang2007electoral}, with the educational levels of the society \citep{lipset1960social}, or even with more market related features, such as the fraction of single-bidder contracts between a government agency and a private company \citep{wachs2021corruption}. Even when this regression-based model approach has many advantages, such as a straightforward implementation, it may also bring some issues, such as the strong underlying hypothesis of a linear dependence between variables, and the emergence of ambiguous causal relationships that may lead to flawed conclusions \citep{depken2006fiscal, loftus1996psychology}. On the other hand, the predictive power of regression models has proved to be poor compared to other modern techniques such as machine learning \citep{delen2012comparative, delen2005predicting, seifert2004data}.

Another common approach to try to characterize and prevent corruption in public procurement is building risk factors from contract data. The Center for Research on Corruption in Budapest has curated a list of the most relevant indicators of corruption risk in public procurement, and they have studied the efficiency of these risk factors \citep{wachs2019social,fazekas2020corruption, david2020grand,fazekas2013anatomy,fazekas2013corruption,fazekas2016objective,
wachs2021corruption}. 
A usual technique analysis is to use linear, or multi-linear, 
regression models to find the best predictors of corrupt practices 
%\citep{charron2017careers,falcon2022practices}. 
In \cite{falcon2022practices} however, we tested four risk factors related to M\'exico's public procurement, finding that isolated risk factors (or even linear combinations of them) result in very poor predictors of corruption. Nevertheless, since the phenomenon of corruption is highly complex, perhaps a better way to predict it would be through non-linear functions of the risk factors.

\subsection{The Predicting power of Machine learning techniques}
%el chido es machine learning
%qué se ha hecho con machine learning en este sentido
%\textcolor{red}{Machine Learning is the field of study that allows computers to learn without being explicitly programmed (Arthur samuel, 1959). Tom Mithell provides a more specific definition \citep{mitchell1997}:

%\textit{A computer program is said to learn from experience E with respect to some task T and some performance measure P, if its performance on T, as measured by P, improves with experience E}. 
%NO ME GUSTA}

The recent development of data modeling algorithms, improvements in the speed and memory capacity of computers, and a large amount of publicly available data, have boosted the machine learning research to develop algorithms with improved detection and prediction capabilities \citep{sharda2022business}. These techniques have become widespread over the last years to study, detect and predict patterns embedded in data, with applications in different areas of science and engineering, such as spam filtering,  tumor detection,  identification of credit card fraud or forecasting a company's revenues \citep{alpaydin2020,geron2019}.    

The use of machine learning tools in understanding, classifying, and predicting corruption and governmental fraud has been gaining popularity in the scientific community in the last few years \citep{stockemer2018internet, sun2019mapping, tang2019effects, lima2020predicting, zumaya2021identifying, li2020detection,ash2021machine,rabuzin2019prediction}. 
%According to \citeauthor{de2019and}, \lq\lq {\it there is a growing interest for studies involving artificial intelligence in the public sector}''. 
Artificial intelligence has contributed to untangling fraud and risk issues in many sectors related to economic affairs by working with massive data sets generated inside government institutions. For example, \citet{zumaya2021identifying} used random forest and artificial neural networks to analyze more than 80 M data from taxpayers in M\'exico to identify shell companies. In \cite{ash2021machine} a gradient boosted classifier of an ensemble of decision trees is implemented to detect and predict the probability of corruption in a municipality, taking into account the data from budget accounts of the Brazilian government. Finally, \citet{rabuzin2019prediction} used text mining techniques to detect risk factors inside public contracts and developed a machine learning model to recognize suspicious tenders in Croatia. 

This paper aims to use machine learning techniques on a dataset of Mexican public procurement contracts to identify and predict those with a high probability of being corrupt, but also to study the contract features that are more relevant in identifying corrupt contracts.
\begin{ThreePartTable}
\begin{TableNotes}
\footnotesize
\item [a] The Mexican Economic Secretariat proposed this size classification based on the amount of resources produced by the company and its number of employees \citep{EstratSE}.
\item [b] Since most of the Spending is reported in mexican currency (MXN), we converted the amounts to USD PPP using the equivalences given in \cite{oecdppp}.
\end{TableNotes}
\begin{longtable*}[htb!]{llll}
\toprule
\hline
 \bf Type & \bf Variable & \bf Short name & \bf Detail\\
 \hline
\endfirsthead

\toprule
\hline
 \bf Type & \bf Variable & \bf Short name & \bf Detail\\
 \hline
\endhead

\bottomrule
\endfoot

\bottomrule
\insertTableNotes  % tell LaTeX where to insert the contents of "TableNotes"
\endlastfoot

i) & Government Order &
  GO &
  \begin{tabular}[c]{@{}l@{}}{\bf Government agency (buyer) level:}\\ APF -  Federal level \\ GE - State level\\ GM - Municipal level %\\ (This variable provides information regarding at which level\\ of government the contract was assigned, as different orders\\ have different operation rules regarding\\ duration, thresholds, etc; and different levels of scrutiny) 
  \end{tabular}   \\
  \hline
  &
Procedure Character &
  PC &
  \begin{tabular}[c]{@{}l@{}} {\bf Legal framework in which the contract was made:}\\N - National\\ I - International\\ ITLC - International under the North American\\ Free Trade Agreement (NAFTA) %\\ (This variable provides information of the legal framework under which\\ the contract was assigned)
  \end{tabular} \\
  \hline
&
Contract Type &
  CT &
  \begin{tabular}[c]{@{}l@{}}{\bf Services or commodities contracted:} \\ OP - Public work \\ S - Services\\ ADQ - Acquisitions\\ AR - Leases\\ SLAOP - Special public work%\\ (This variable provides a rough description of the market sector\\ of the contract, as again,  different sectors might, in principle,\\ have different regulations) 
  \end{tabular} \\
  \hline
  &
Procedure Type &
  PT &
  \begin{tabular}[c]{@{}l@{}} {\bf Procedure by which the supplier won the contract:}\\ AD - Single-bidder\\ LP - Open public contest\\ I3P - Contest between only three suppliers %\\ (Describes how the contract was assigned. We recall that,\\ in principle, single-bidder contracts can be carried \\out only under \lq\lq special" circumstances)  
  \end{tabular} \\
  \hline
&
Size &
  S &
\begin{tabular}[c]{@{}l@{}}{\bf Size of supplier:} \tnote{a}
\\
MIC - Micro-supplier\\ PEQ - Small-supplier\\ MED - Medium-supplier\\ NOM - Large-supplier\\ NA - Supplier without assigned size%\\ (This variable is a rough indicator of the size of the supplier,\\  which can serve as a proxy  of the capacity of the supplier\\ to carry out the project, etc) 
\end{tabular} \\
  \hline
&
Beginning week &
  N/A & \begin{tabular}[c]{@{}l@{}}
  {\bf Week of the year in which the contract began} %\\ (This variable intends to gauge whether the activity\\ is correlated to the budget calendar) 
  \end{tabular}\\
  \hline
&
Ending week &
  N/A & \begin{tabular}[c]{@{}l@{}}
  {\bf Week of the year in which the contract ended}
  \end{tabular}\\
  \hline
&
Ending-Beginning weeks &
  EBWeeks &
  {\bf Weeks that the contract lasted} \\
  \hline
  &
Spending &
  N/A &
  {\bf Amount of money spent by the buyer (in USD PPP)} \tnote{b} 
  \\ \hline
  
ii) &
  \begin{tabular}[c]{@{}l@{}}
  Number of Contracts
  by a buyer
  \end{tabular}
  &
  T.Cont &
  \begin{tabular}[c]{@{}l@{}}{\bf Number of total contracts given to each supplier} \end{tabular} \\
  \hline
 &
  Spending by a buyer &
  T.Spending &
  \begin{tabular}[c]{@{}l@{}}{\bf Amount of money given to each supplier}  \end{tabular} \\ 
  \hline
   &
  Single-bidder contracts by  a buyer &
  T.AD &
  \begin{tabular}[c]{@{}l@{}}{\bf Number of total contracts given to each supplier} \\ {\bf by single-bidder}  \end{tabular} \\ 
  \hline
     &
  Active weeks &
  N/A &
  \begin{tabular}[c]{@{}l@{}}{\bf Number of weeks in which a supplier was assigned} \\ {\bf contracts by a buyer in each year}  \end{tabular} \\ 
  \hline
iii) &
  \begin{tabular}[c]{@{}l@{}}
  Maximum number of Contracts by a buyer
  \end{tabular}
  &
  T.Cont.Max &
  \begin{tabular}[c]{@{}l@{}}{\bf Maximum number of the contracts awarded by} \\ {\bf a buyer to a supplier}\\ $\max_j{(T.Cont_j)}$ where $j$ represents all the suppliers contracted\\ by a buyer,and $T.Cont$ the number of total contracts\\ given to each supplier \end{tabular} \\
  \hline
 &
  Maximum Spending by a buyer &
  T.Spending.Max &
  \begin{tabular}[c]{@{}l@{}}{\bf Maximum amount of money spent by a buyer}\\ {\bf in contracts with a supplier}\\ $\max_j{(T.Spending_j)}$ where $j$ represents all the suppliers\\ contracted by  the buyer, and $T.Spending$ the amount of money\\ given to each supplier\end{tabular}\\ \hline
  iv)
 &
  Fraction of single-bidder contracts &
  RAD &
  \begin{tabular}[c]{@{}l@{}}{\bf Fraction of single-bidder contracts awarded}\\ {\bf by a buyer to a supplier:} $T.AD/T.Cont$\end{tabular} \\ \hline
&
  Favoritism &
  Fav &
  \begin{tabular}[c]{@{}l@{}}{\bf Favoritism of a buyer for a supplier as defined in} \\ \cite{IMCOmapeando}
  \\ $(0.33)*(T.Cont/T.Cont.Max)$ + \\ $(0.66)*(T.Spending/T.Spending.Max)$\end{tabular} \\
  \hline
 &
  Contracts per active week &
  CPW &
  \begin{tabular}[c]{@{}l@{}}{\bf Number of contracts a supplier won} \\ {\bf with the same buyer per active week}\\ $T.Cont/ActiveWeeks$\end{tabular} \\
  \hline
 &
  Spending per active week &
  SPW &
  \begin{tabular}[c]{@{}l@{}}{\bf Amount of money spent by a buyer in contracts} \\ {\bf with the same supplier per active week}\\ $T.Spending/ActiveWeeks$\end{tabular} \\
  \hline \hline
    \caption{{\it Predictor variables:} {Type i):} Set of the variables chosen to describe the features of each contract from the source list of public contracts \citep{compranet}. {Type ii):} Set of variables to describe the relationship buyer/supplier. {Type iii):} Set of variables to describe features of the buyers. {Type iv):} These variables are approximate versions of the risk factors proposed in \cite{fazekas2013anatomy,fazekas2013corruption, IMCOmapeando}, see text.}
\label{datoscontratos}\\
 \end{longtable*}
\end{ThreePartTable}

\section{Methods}
\label{sec:methods}

\subsection{Data Preparation}
\label{sec:data_prep}

As mentioned above, the data set used in this investigation is essentially the same as that analyzed in \cite{falcon2022practices}. The only difference is that the previous data used dummy variables for each categorical variable, while here we keep categorical variables as such. In this section we give a short description of the data and variables considered by the machine learning model.

The list of public contracts was taken from the electronic Mexican system of public governmental information on public procurement {\it CompraNet} \citep{compranet}, and it covers all the public contracts from 2013 to 2020. The {\it CompraNet} system is operated by an administrative unit designed by the Mexican federal agency for budget and public debt (SHCP for \lq\lq Secretar\'ia de Hacienda y Cr\'edito P\'ublico''). The registry of contracts in this electronic system is mandatory for all those that operate with resources from the federal budget \footnote{These include federal, state, and municipal agencies. It should also be mentioned that we have no way of knowing how complete the list is, nor whether omissions are more frequent regarding contracts in one government level or another, as well as the possibility of omissions regarding \lq\lq sensitive" contracts, as could be military spending. Finally, we remark that the list only includes contracts between government agencies and {\it private} suppliers. Government agencies that act as suppliers to other government agencies are exempt from reporting in {\it CompraNet}.}. The contracts on this list have a specific set of variables or entries that describe them. The particular entries we use in this work are shown in Table \ref{datoscontratos} - Type i).

The original data lists consisted in 1.6 M of contracts. These records were curated to standardize the information available in each of them. We homogenized all the string variables to avoid issues with special characters or spacing \footnote{By law all registries in {\it CompraNet} should be made using the fiscal name of both the agency and company. This ensures that the data contains unique identifiers of buyers and suppliers.}, we also deleted all the entries in which an important variable was omitted (for example, the buyer's name, the amount spent, the type of procedure, etc. ). The contracts with this kind of problems were approximately 60 K (representing $\sim$4\% of the data). Thus, from the 1.6 M contracts available in the original data lists, we consider 1.54 M. As mentioned above, from these 1.54 M of records, we took only the variables shown in Table \ref{datoscontratos} - Type i). The original set includes other variables that are not of interest for this work, such as the name of the buyers' legal representative, or the suppliers' webpage. Hereinafter, we will refer to this curated list of 1.54 M of contracts as the {\it source list}.

We use the information given in the source list to build a few extra variables. We constructed three variables related to the relationship between buyers and suppliers (Table \ref{datoscontratos} - Type ii)), and two variables that are descriptive of the buyers (Table \ref{datoscontratos} - Type iii)).  Unfortunately, the available data in the source list is not detailed enough to evaluate exactly the risk factors developed in \cite{fazekas2013anatomy,fazekas2013corruption, IMCOmapeando}, thus, we propose approximate versions of the factors that give similar information about the features of the relationship between the buyers and suppliers, and we refer to these variables as Type iv) (See Table \ref{datoscontratos}). %the total number of contracts given to a supplier from a buyer ({\bf T.Cont}), the total amount spent by the buyer in the same supplier ({\bf T.Spending}), and the total single-bidder contracts between a buyer and a supplier (\bf T.AD) (Table \ref{datoscontratos} - Type ii)). We also build two variables that are descriptive of the buyers: the maximum number of contracts assigned by the buyer to the same supplier, and the maximum total amount spent by the buyer with the same supplier in each year. We call these variables {\bf T.Cont.Max} and {\bf T.Spending.Max}. These variables give us an idea of the budget managed by each buyer and its activity in the public procurement market (Table \ref{datoscontratos} - Type iii)). Unfortunately, the available data in the source list is not detailed enough to evaluate exactly the risk factors developed in \cite{fazekas2013anatomy,fazekas2013corruption, IMCOmapeando}, thus, we propose approximate versions of the factors that give similar information about the features of the relationship between the buyers and suppliers, and we refer to these variables as Type iv) (See Table \ref{datoscontratos}).\\

Thus, considering the information available in the source list, and the variables related to buyer/supplier features, buyers' features and risk factors, there are four types of items in the final data set: 

\begin{enumerate}
    \item [i)] Items that describe the features of the contracts. For example, the procedure by which the supplier won the contract (single-bidder, public contest, etc.), the amount allocated to the contract, and the week in which the contract began.
    \item [ii)] Items that describe the  features of the relationship buyer/supplier. These include the amount spent by a buyer with a supplier, or the number of contracts carried out between a buyer and the same supplier per year. 
    \item [iii)] Items that describe the  features of the buyers. These include the maximum amount spent by a buyer with a supplier, or the maximum number of contracts carried out by a buyer with a single supplier per year. 
    \item [iv)] Items that give information about the relationship between buyer/supplier and have been used as corruption risk factors \citep{fazekas2013corruption,wachs2021corruption,IMCOmapeando}. Examples of these are the fraction of single-bidder contracts awarded by a buyer to supplier, or the favoritism of a buyer for a supplier.
\end{enumerate}

Thus, we end up with a list of contracts where each one is described by 19 variables, 5 of which are categorical and 14 are numerical. The statistical information of the numerical variables is shown in Table \ref{tab:datastats}.

\begin{table}[htb!]
\begin{tabular}{llllll}
\hline \hline
{\bf Type} & {\bf Variable}       & {\bf Mean}                & {\bf SD} & {\bf 1st Qu.}             & {\bf 3rd Qu.}             \\
\hline \\
i)   & Beginning week & 25.14               & 14.73   & 12.00               & 38.00               \\
     & Ending week    & 35.24               & 16.39   & 22.00               & 52.00               \\
     & EBWeeks        & 15.54               & 22.68    & 1.00                & 23.00               \\
     & Spending       & 3.3 $\times 10^{5}$ & 1.3 $\times 10^{6}$    & 4.7 $\times 10^{3}$ & 5.5 $\times 10^{4}$ \\
     \hline \\
ii)  & T.Cont         & 117.20               & 334.39    & 1.00                & 35.00               \\
     & T.Spending     & 1.3 $\times 10^{7}$ & 1.1 $\times 10^{7}$   & 2.7 $\times 10^{4}$ & 1.6 $\times 10^{6}$ \\
     & T.AD           & 112.10               & 328.62   & 1.00                & 29.00               \\
     & Active weeks   & 12.74               & 17.09   & 1.00                & 18.00               \\
     \hline\\
iii) & T.Cont.Max     & 534.5               & 818.84   & 11.0                & 650.0               \\
     & T.Spending.Max & 4.9 $\times 10^{8}$ & 7.7 $\times 10^{7}$    & 7.7 $\times 10^{7}$ & 8.0 $\times 10^{8}$ \\
     \hline \\
iv)  & RAD            & 0.74                & 0.38   & 0.53                & 1.00                \\
     & FAV            & 0.12                & 0.17   & 0.018               & 0.15                \\
     & CPW            & 3.5                 & 6.72   & 1.0                 & 2.0                 \\
     & SPW            & 7.4 $\times 10^{5}$ & 2.4 $\times 10^{6}$   & 1.2 $\times 10^{4}$ & 1.9 $\times 10^{5}$\\
     \hline\hline
\end{tabular}

\caption{Descriptive statistics for all the numerical predictor variables (types ii, iii, and iv). The table shows the mean, the standard deviation (SD), and the 1st and 3rd quartiles.}
\label{tab:datastats}
\end{table}

Since 2013 M\'exico's government has collected a list of companies that have been caught participating in corrupt activities such as: selling fake receipts to buyers who use them to avoid taxes or to cover acts of embezzlement; presenting fake documentation to win a contract; overcharging their services: breaching contract; or diverting resources \footnote{It should be noted that to appear in these lists, these contractors were subject to a legal investigation, and some of the contractors are appealing the decision. Thus, these lists may change slightly as time goes by due to companies winning their appeals, which removes them from the lists, and conclusions of long-lasting investigations may add new companies.}. The list of these allegedly \lq\lq corrupt" companies is made by merging the lists available on the site of the Mexican tax agency (SAT for \lq\lq Secretar\'ia de Administraci\'on Tributaria'') \citep{EFOS} \footnote{For this study we use the lists of both definitive and presumed corrupt companies}, and of the official Mexican open data site \citep{PCS}. The only variable of interest in this list of corrupt companies is the supplier's name, which we use to identify the contracts in which they participated.

\begin{figure*}[htb!]
    \begin{center}
        \includegraphics[width=0.9\textwidth]{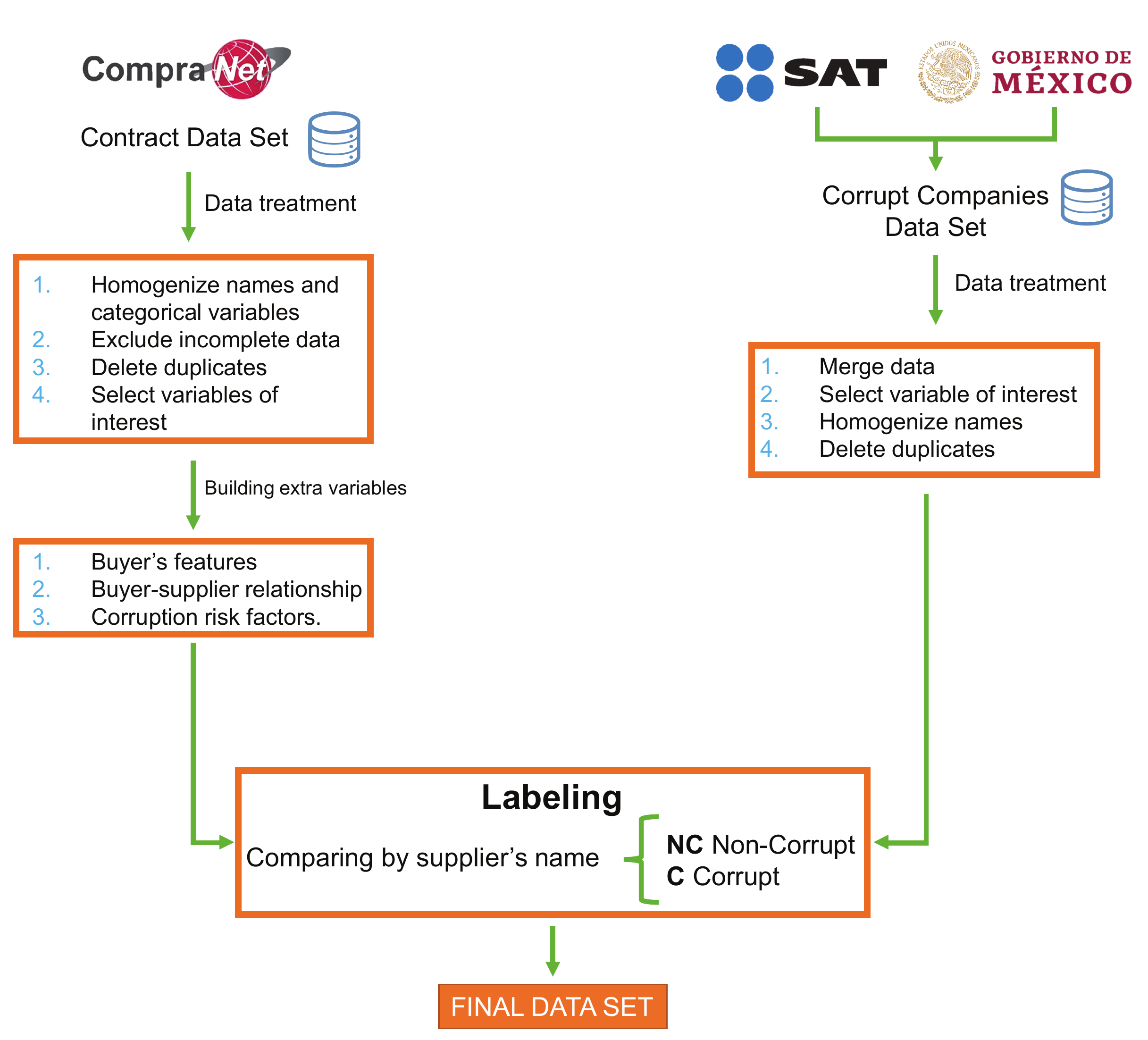}
    \end{center}
    \caption{Graphical depiction of data preparation and processing. Broadly: The contract dataset (left branch) was obtained from the open site {\it CompraNet}. The dataset was treated following, in order, the steps shown in the first box, "Data treatment", to finally select the variables of interest (Type i)). From this new curated data, we build some extra variables shown in the second box, "Building extra variables" (Types ii), iii) and iv)). On the other hand, the corrupt companies' dataset (right branch) was obtained by merging the lists available on the Mexican Tax Agency's site and on the Official Mexican Open Data's site and was curated following, in order, the steps shown in the first box "Data treatment". Then, we labeled each contract in the contract dataset by comparing its supplier's name with the list of corrupt companies. If the supplier is in the corrupt companies' dataset, then we assigned the Corrupt ($C$) label; otherwise, we gave the Non-Corrupt label ($NC$). This labeled dataset is our final dataset.}
    \label{fig:Diag_Datos}
\end{figure*}

Next, we labeled those public procurement contracts won by the corrupt companies. Since these companies have gone through a process to be classified as corrupt, we assume that they may be suspect of having incurred in corrupt behavior in all the contracts they participated in, independently of whether the contracts occurred before or after the company was labeled. Thus, all the contracts in which the supplier is a corrupt company are classified as corrupt, and we assign the corresponding label $C$ (for Corrupt) to them. Following the principle of presumption of innocence, we label as $NC$ (for Non-Corrupt) all the other contracts in the source list in which the supplier is free of official corruption charges. Thus, we have two classes of contracts labeled $C$ and $NC$, respectively. Of course, we are aware that it is very likely that corrupt contracts went undetected and ended up in our $NC$ class; however, we expect that they will have little statistical weight in this class that represents the vast majority of the contracts. Figure \ref{fig:Diag_Datos} shows a simple diagram that summarizes the whole data process described above.

All datasets mentioned in this section, and all the tools needed to reproduce the results shown here are available on \cite{zenodo22}.

\subsection{Training, Calibration and Test Data Sets.}
\label{sec:split_datasets}

To assess the performance of any classification algorithm on new cases, it is a common practice to split the dataset into Training and Testing Data Sets \citep{alpaydin2020,geron2019,mitchell1997}. As the names suggest, we train the algorithm using the Training Set and evaluate its performance using the Test Set. The model we propose in this paper, described in detail in section \ref{sec:model}, is conformed of three main steps: training, calibration and testing. We use the calibration step to control the proportion of true positives and false positives, meaning the fraction of actual non-corrupt and corrupt contracts classified as $NC$. These values are the True Positive Rate (TPR) and False Positive Rate (FPR). In the rest of this paper, we consider $NC$ as the positive class. Following the same principle to split the data in Training and Test Data Sets, we applied an additional split to calibrate the TPR and FPR with an independent dataset different from the Training and Test Sets. Thus, we split the entire dataset into Training, Calibration and Test Sets. %We show the details of this partition in Table \ref{tab:SplitSets}. 

\subsection{Imbalanced data}
\label{sec:imbalanced_data}

A data set is imbalanced when one class is significantly more represented than the others, {\it i.e.}, when the number of instances of one class is significantly larger than the instances in other classes. As shown in the Table \ref{tab:totalclases}, the public procurement dataset contains 1,506,892 (97.8\%) $NC$ versus 33,494 (2.2\%) $C$ instances (contracts), with an $NC:C$ ratio of 45:1. 

\begin{table}[htb!]
\begin{tabular}{l@{\hspace{1em}}l@{\hspace{1em}}l}
\hline\hline
\textbf{Contract Type} & \textbf{Total} & \textbf{Percentage} \\
\hline
Non-Corrupt ($NC$)       & 1,506,892      & 97.8                \\
Corrupt ($C$)            & 33,494         & 2.2                 \\
Total                  & 1,540,386      & 100\\
\hline\hline
\end{tabular}
\caption{Total number of contracts in each class. The ratio of $NC$ vs. $C$ contracts is roughly 45:1.}
\label{tab:totalclases}
\end{table}

In every machine learning classification algorithm, imbalanced data is likely to produce undesirable effects, such as over fitting for the most represented class, poor performance classifying the underrepresented data and much lower performance in the Test than in the Training Data \citep{geron2019,Japkowicz2002, Khalilia2011,kuhn2013}. 

Different techniques mitigate the adverse effects of class imbalance, including oversampling, undersampling, boosting, repeated random sub-sampling or combinations of them \citep{Japkowicz2002,quinlan1996}. In the present paper, we implemented Repeated Random Sub-sampling as described in  \citep{Khalilia2011} to contend with our dataset's imbalanced $C$ and $NC$ classes. With this approach, we generate $|NC|/|C|\approx 45$ which are the maximum number of random sub-samples without replacement that we can take from the Training Data, where $|NC|$ and $|C|$ are the amount of non-corrupt and corrupt contracts, respectively. Then, each sub-sample comprises the $|C|$ corrupt contracts and an equal amount of non-corrupt contracts randomly selected without replacement. The complete learning model, described in section \ref{sec:model} trains one classifier for each balanced sub-sample and implements a \lq\lq voting'' model to determine the final class membership.

\subsection{Feature Selection.}

\begin{figure*}[htb!]
    \begin{center}
        \includegraphics[width=0.95\textwidth]{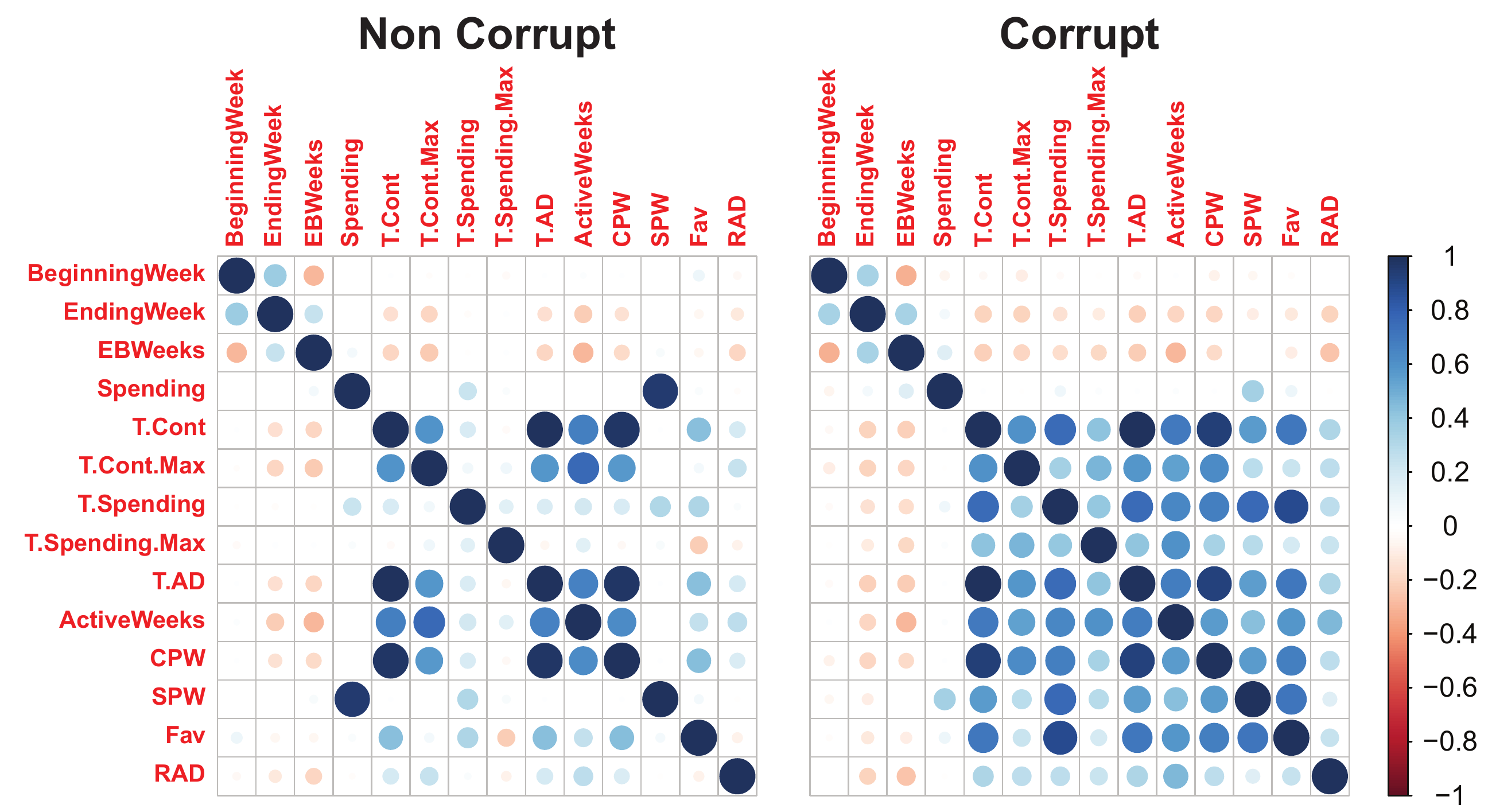}
    \end{center}
    \caption{Correlation matrices for Non-Corrupt and Corrupt contracts. Color code indicates the strength of the correlation. Most of the correlations in non-corrupt contracts are modified in the corrupt dataset. Also, variables {\bf T.Cont} to {\bf RAD} are more positively correlated in the case of corrupt contracts. Variables {\bf BeginningWeek} to {\bf EBWeeks} are weakly anti-correlated with the rest of the features.}
    \label{fig:Correlations}
\end{figure*}

As previously mentioned, 19 features describe each record in the final dataset: 5 categorical and 14 numerical. Also, each record is labeled as Corrupt ($C$) or Non-Corrupt ($NC$) according to the labeling described previously in section \ref{sec:data_prep}. Feature selection is a common step in training machine learning systems. Redundant features may negatively affect the learning performance, predictive accuracy and comprehensibility of learned results \citep{kuhn2013}. To avoid redundant variables, we calculated the linear Pearson's correlation between the 14 numerical features in the dataset. Given the high imbalance between $NC$ and $C$ contracts, we implemented this process separately for both datasets and looked for common correlated variables. We found that the correlation coefficient of every pair of variables differs by more than 5\% between $C$ and $NC$, as shown in Figure \ref{fig:Correlations}. For this reason, we integrated all 19 features as possible predictors for our machine learning model.

\begin{figure*}[htb!]
    \begin{center}
        \includegraphics[width=0.98\textwidth]{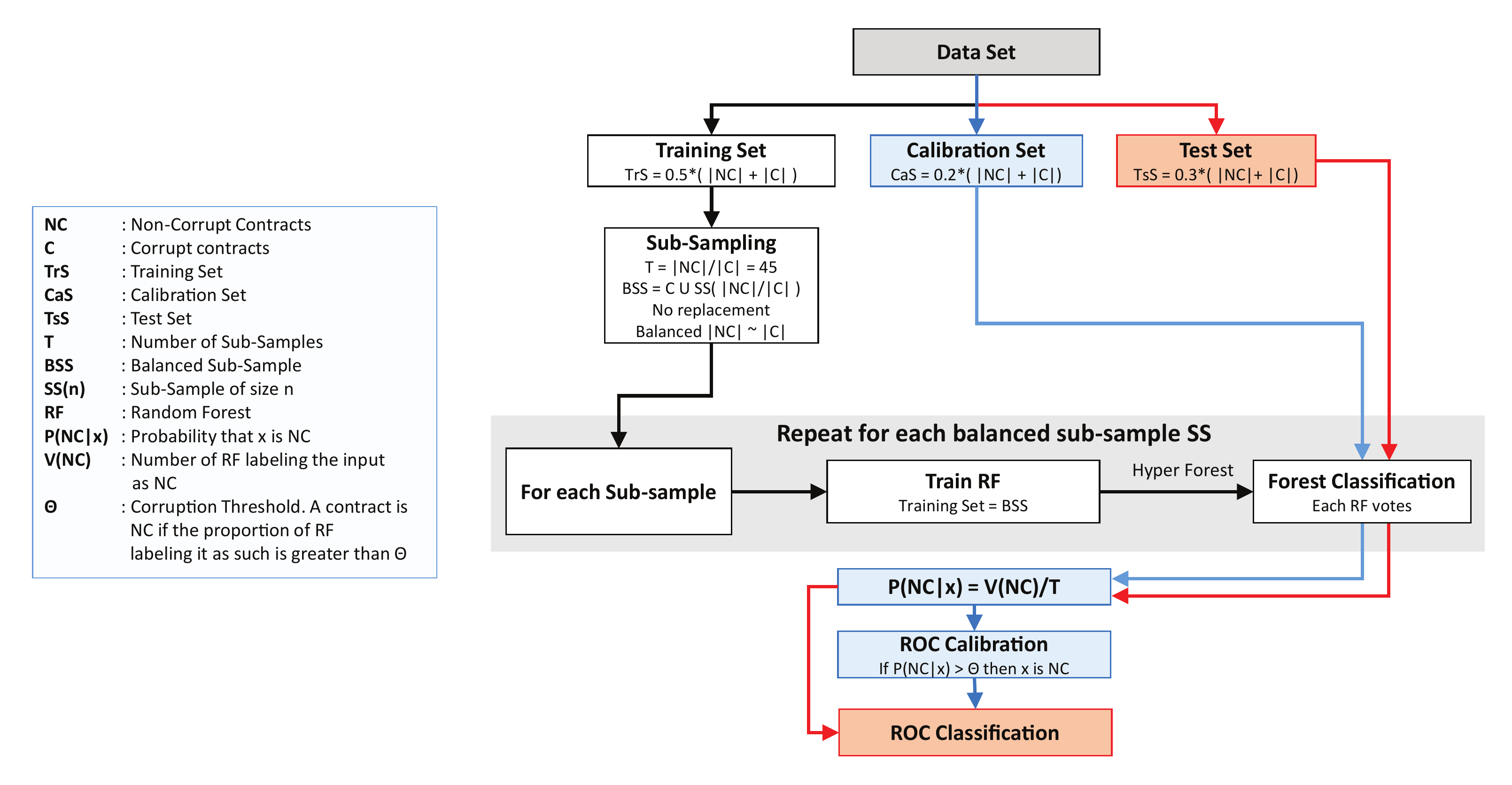}
    \end{center}
    \caption{Flowchart of the machine learning model for classifying public procurement contracts. The data is partitioned into Training, Calibration and Test Sets. The Training Set is split to produce a set of balanced sub-samples containing an equal amount of Non-Corrupt ($NC$) and Corrupt ($C$) contracts. A Random Forest (RF) is then trained with each sub-sample. All RFs vote to classify a new input contract $x$, and the voting result produces the probability ${\mbox P}(NC|x)$ that $x$ is Non-Corrupt. A ROC curve is used with the Calibration Set to select the threshold $\theta$ such that a contract is classified as $NC$ if ${\mbox P}(NC) > \theta$. The white, blue and red boxes indicate sub-processes related to the training, calibration or test stages.}
    \label{fig:Diag_ML}
\end{figure*}

\subsection{Random Forest}
\label{sec:random_forest}

The backbone machine learning algorithm we use in this paper is the ensemble learner Random Forest (RF) \citep{breiman2001random,breiman2017}. This algorithm creates multiple decision trees for classification/regression problems. Decision trees approximate discrete-valued functions by a set of hierarchical concatenated decisions represented by a tree \citep{breiman2017,mitchell1997}. Each decision tree in the RF classifier is trained with a random subset of the data (bootstrapping). It starts with a single root node and selects a feature to split its training sub-set into two or more \lq\lq child" nodes. This split partitions the data into smaller, more homogeneous groups. In this context, homogeneity means that the nodes resulting from the split contain a higher proportion of elements of one class than the precursor nodes. The splitting process is repeated for all the nodes until all elements in the leaves are completely homogeneous; the grown trees are not pruned. At every node, a small number of features is selected randomly without replacement for the split.
Each trained decision tree in the RF classifies new instances by moving them down the tree, starting at the root and finishing on some leaf, which provides the label to classify the instances. Every node tests a feature of the instance in question, the possible values of the test correspond to branches to other nodes further down the tree, where other tests will be applied. This process is repeated until it reaches a \lq\lq leaf" corresponding to the classification of the instance \citep{alpaydin2020, breiman2017, kuhn2013,mitchell1997}. In a RF, a new instance is classified in this way by each of the trees, and the majority determines the new instances's final classification \citep{breiman2001random,geron2019,kuhn2013}.\\

RF models are widely used for many classification and regression problems \citep{Chen2012,Gislason2006,JiongZhang2008,Khalilia2011,lima2020predicting,Qi2012}. They can handle high dimensional data, work naturally with categorical and numerical features, integrate many classifiers (trees) for the ensemble and estimate the importance of the features used for the classification.

\subsection{Importance of the Variables}
\label{sec:importancevariable}

One of the essential features of RF is that it allows us to evaluate the relevance of a variable as a predictor by calculating its importance. Importance of the variable measures the relationship between a variable and the classification result \citep{breiman2001random,Khalilia2011}. A variable $X_i$ is important for predicting the class $Y$ if the prediction error increases by breaking the relationship between $X_i$ and $Y$. A RF can compute this measurement in various ways \citep{breiman2001random,kuhn2013}: permutation, z-score and Gini importance. We focus on the permutation importance described in \cite{breiman2001random}. Broadly, Breiman proposes to take a first measure of the average accuracy in the \lq\lq out-of-bag samples" ({\it i.e.} those that did not participate in the training) of each tree. Then apply a random permutation on the values of each variable $X_i$ in the out-of-bag samples, measure the accuracy again and record the decrease in the accuracy under the permutation. Each tree carries out these calculations during the RF construction. Finally, the decrease in the accuracy due to the permutation is averaged over all the trees in the RF and is reported as the importance of the variable $X_i$. 
The random permutation breaks the relationship between $X_i$ and the predicted variable $Y$. It also breaks the link between $X_i$ and other covariates, but leaves the distribution of values of $X_i$ intact. Variables $X_i$ with larger values of the mean decrease in accuracy suggest a strong link between $X_i$ and $Y$. In contrast, smaller values indicate a weaker association and therefore are less critical as predictors.

\subsection{Receiver Operating Characteristic (ROC) curve}
\label{sec:roc}

A ROC curve is a tool used to evaluate and calibrate the performance of a binary classifier \citep{alpaydin2020,Bradley1997,geron2019}. Let us consider that a classifier returns ${\mbox P}(C_1|x)$, the probability that the input data $x$ belongs to class $C_1$ and ${\mbox P}(C_2|x) = 1 - {\mbox P}(C_1|x)$, the probability that $x$ belongs to class $C_2$. For definiteness, we choose $C_1$ as the positive class. Now we choose a threshold $\theta$, such that the input $x$ is classified as $C_1$ if ${\mbox P}(C_1|x) > \theta$. If $\theta$ is close to 1, very few instances are classified as $C_1$, decreasing both the number of false positive results (FPR) and true positive results (TPR). Conversely, decreasing $\theta$ will increase the number of TPR, at the risk of increasing the FPR \citep{alpaydin2020,geron2019}. The ROC curve plots the TPR against the FPR for $\theta$ in the range $[0,1]$. An ideal classifier has a TPR=1 and a FPR=0. Thus the classifier can be considered better the closer the ROC curve passes to the coordinates $(0,1)$ in the upper left corner of the ROC space \citep{alpaydin2020}. The point of the ROC curve closest to $(0,1)$ corresponds to the value of $\theta$ that maximizes the trade-off between TPR and FPR. In contrast, the diagonal in the ROC space corresponds to the ROC curve of a purely random classifier that makes as many true decisions as false ones.

The ROC curve allows for visual analysis: one way to numerically compare the performance of two classifiers is by calculating the Area Under the ROC Curve (AUC) \cite{Bradley1997}. A perfect classifier has an AUC of 1, a purely random classifier will have an AUC equal to 0.5, and the AUC values of different classifiers can be compared to give a general idea about their performance \citep{alpaydin2020}.

\subsection{Recursive Feature Elimination}
\label{sec:RFE}

Feature selection aims to remove non-relevant or redundant features as model predictors. Those variables may introduce uncertainty, noise, and difficulty interpreting the results. Also, they can negatively affect the performance of a classification algorithm \citep{kuhn2013}. 

Recursive Feature Elimination (RFE) is a variable selection algorithm that matches naturally with the importance of the variables of the RF to remove non-informative or redundant features from the model. It works as a backward selection algorithm: The initial model contains the entire set of features, which are then removed iteratively to determine those that are not contributing to the model's performance. Then the model is rebuilt with the remaining variables. In RF classifiers, once the entire model is created, the importance of the variables is calculated (as described in section \ref{sec:importancevariable}). At each stage of RFE, the least important variable is eliminated before rebuilding the model. Once the new model is trained, the accuracy is estimated for that model. This process continues until it reaches a stop criterion. In our approach, the process stops when there are no more features to delete. The final set of predictors is the subset of variables with the best accuracy \citep{guyon2003,kuhn2013}.

\section{Machine Learning Model: Repeated Random Sub-Sampling with Random Forest}
\label{sec:model}

As we mentioned earlier, imbalanced data may negatively affect the performance of any classifier, including RF. To mitigate these effects, we combined RF with Repeated Random Sub-sampling to train multiple RF with balanced sub-samples. Figure \ref{fig:Diag_ML} presents a flow diagram of the machine learning classification system we propose in this paper. The procedure is as follows:

\begin{itemize}
    \item After the initial split of the complete unbalanced dataset into Training, Calibration and Test Sets as described in section \ref{sec:split_datasets}, we split the Training Set with Repeated Random Sub-sampling to produce 
    $|NC|/|C|=45$ balanced training sub-samples, each containing the total corrupt ($C$) registers in the training set, and an equal amount of non-corrupt ($NC$) contracts selected randomly without repetition. 
    
    \item We generate a {\it Hyper-Forest} composed of 45 RFs, each trained with a different balanced sub-sample. In this way, the RFs are unbiased regarding the number of $C$ and $NC$ contracts, and each NC contract participates in the training.
    
    \item For new input data $x$, each RF in the Hyper-Forest votes if the input corresponds to a $C$ or $NC$ contract. If we define $V(NC)$ as the number of RFs voting $x$ as $NC$, we calculate the probability that $x$ is $NC$, ${\mbox P}(NC|x) = V(NC)/T$, were $T$ is the total number RFs. 
    
     \item To finally classify an input contract as $C$ or $NC$, we define a threshold $\theta$ representing the proportion of voters necessary to consider the input $x$ as $NC$. Thus, the new input is classified as non-corrupt if ${\mbox P}(NC|x) > \theta$. Different values of $\theta$ affect the True Positive Rate (TPR) and False Positive Rate (FPR): small values of $\theta$ increase both TPR and FPR, whereas large values of $\theta$ decrease the FPR  at the cost of reducing the TPR. We use a Receiver Operating Characteristic (ROC) curve as described in section \ref{sec:roc} to find the value of $\theta$ that gives the best trade-off between the TPR and FPR. To avoid overfitting, we first train the Hyper-Forest with the Training Set and then use the Calibration Set to calculate the best value of $\theta$ as shown in Figure \ref{fig:ROC}.
    
\end{itemize}

The performance of the classifier we propose in this paper is evaluated in the next section.

\section{Results}

We implemented the RF machine learning model proposed in this paper in R, a popular open source statistical programming language, with the help of the \textit{randomForest} \citep{randomforestR} (Random Forest implementation), \textit{mltest} \citep{mltestR} (classification evaluation) and \textit{pROC} \citep{pROCR} (ROC curve analysis) R packages. Each RF was trained with the default parameters of the function \textit{randomForest}\footnote{Number of trees per forest: 500; Number of features selected at each split: $\sqrt{p}$ where $p$ is the number of features; Minimum size of terminal nodes: 1. The rest of the parameters can be consulted in \cite{randomforestR}}. The dataset we used comprises the 33,497
corrupt and 1,506,892 non-corrupt contracts, each
characterized by 19 total features (Table \ref{tab:totalclases}). One half of each of these sets were used as the Training Set (the
proportion and number of contracts in each stage of the model are shown in Table \ref{tab:SplitSets}).
After applying Repeated Random Sub-sampling in the
Training Set, we built 45 balanced sub-samples
with 33,494 elements. Each of the sub-samples contained the 16,747
Corrupt contracts of the Training Set, and an equal amount of Non-Corrupt
contracts chosen at random without repetition.
 
\begin{table}[htb!]
\begin{tabular}{lcccc}
\hline
 \textbf{Data Set} & \textbf{C}  & \textbf{NC} & \textbf{Total} & \textbf{Fraction of Total} \\ \hline
 Training & 16,747 & 753,446 & 770,193 & 0.5 \\
 Calibration & 6,699 & 301,378 & 308,077 &  0.2\\
 Test & 10,048 & 452,068 & 462,116 &  0.3\\ \hline
\end{tabular}
\caption{Number of contracts in the Training, Calibration and Test Data Sets.}
\label{tab:SplitSets}
\end{table}

\begin{figure}[htb!]
    \begin{center}
        \includegraphics[width=0.45\textwidth]{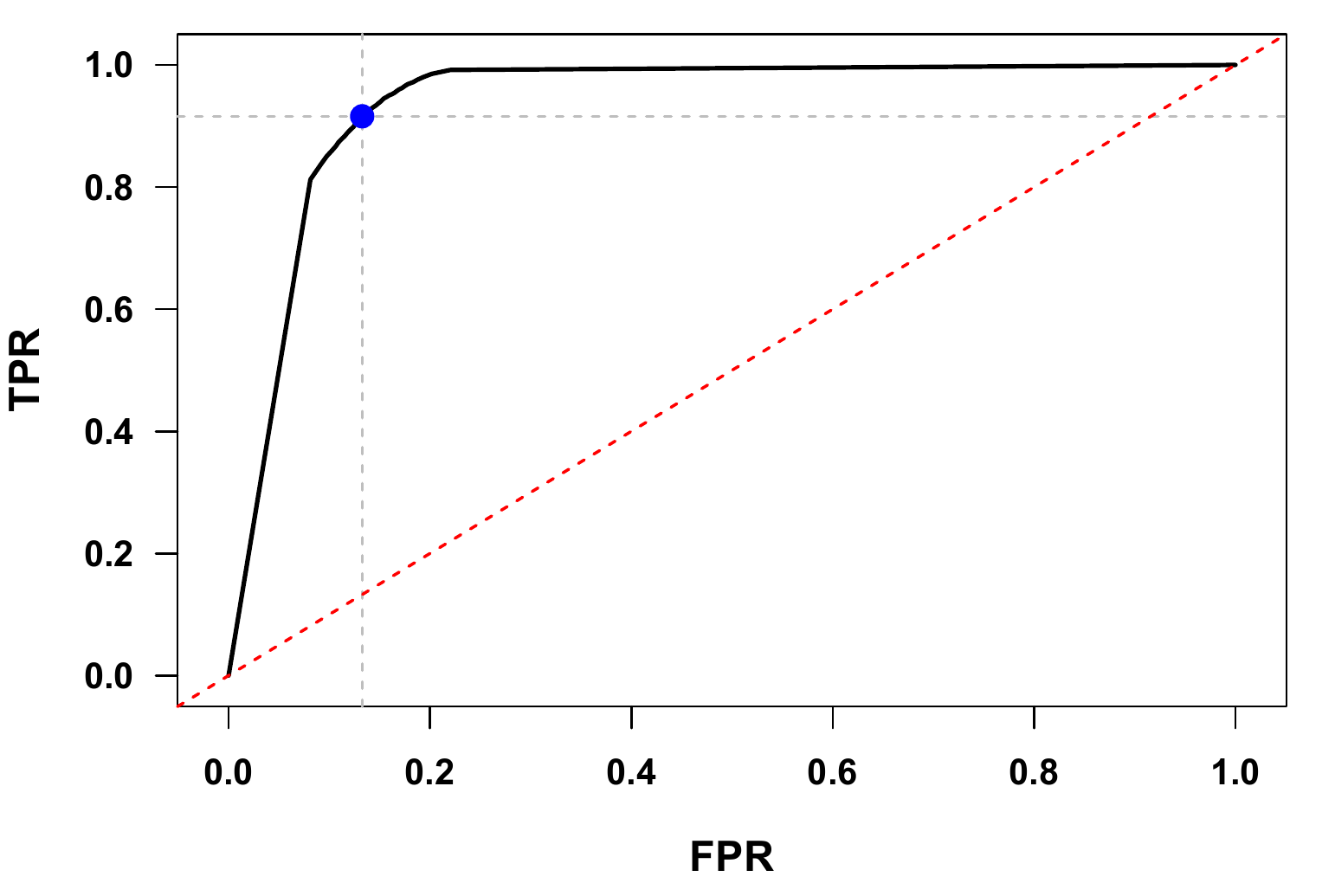}
    \end{center}
    \caption{ROC calibration for the Hyper-Forest classification. The dashed red line indicates theoretical results using a random classifier. The blue point marks the best compromise between TPR=0.92 and FPR=0.13, with an associated threshold $\theta= 0.61$. AUC=0.94.
}
    \label{fig:ROC}
\end{figure}

We constructed the Hyper-Forest consisting of 45 RF, each trained with a different balanced sub-sample. We used a ROC curve and the Calibration set to select the threshold $\theta$ that gives the best trade-off between the TPR and the FPR, as described in sections \ref{sec:roc} and \ref{sec:model}. Figure \ref{fig:ROC} shows the results of the calibration step. The blue point indicates the best compromise between TPR$=0.92$ and  FPR$=0.13$ for $\theta=0.61$. The Area Under the ROC Curve AUC$=0.94$ suggests that the specific value of $\theta$ does not change the TPR and FPR significantly unless $\theta$ is very close to the extreme values $(0,1)$.

\begin{figure}[htb!]
    \begin{center}
        \includegraphics[width=0.45\textwidth]{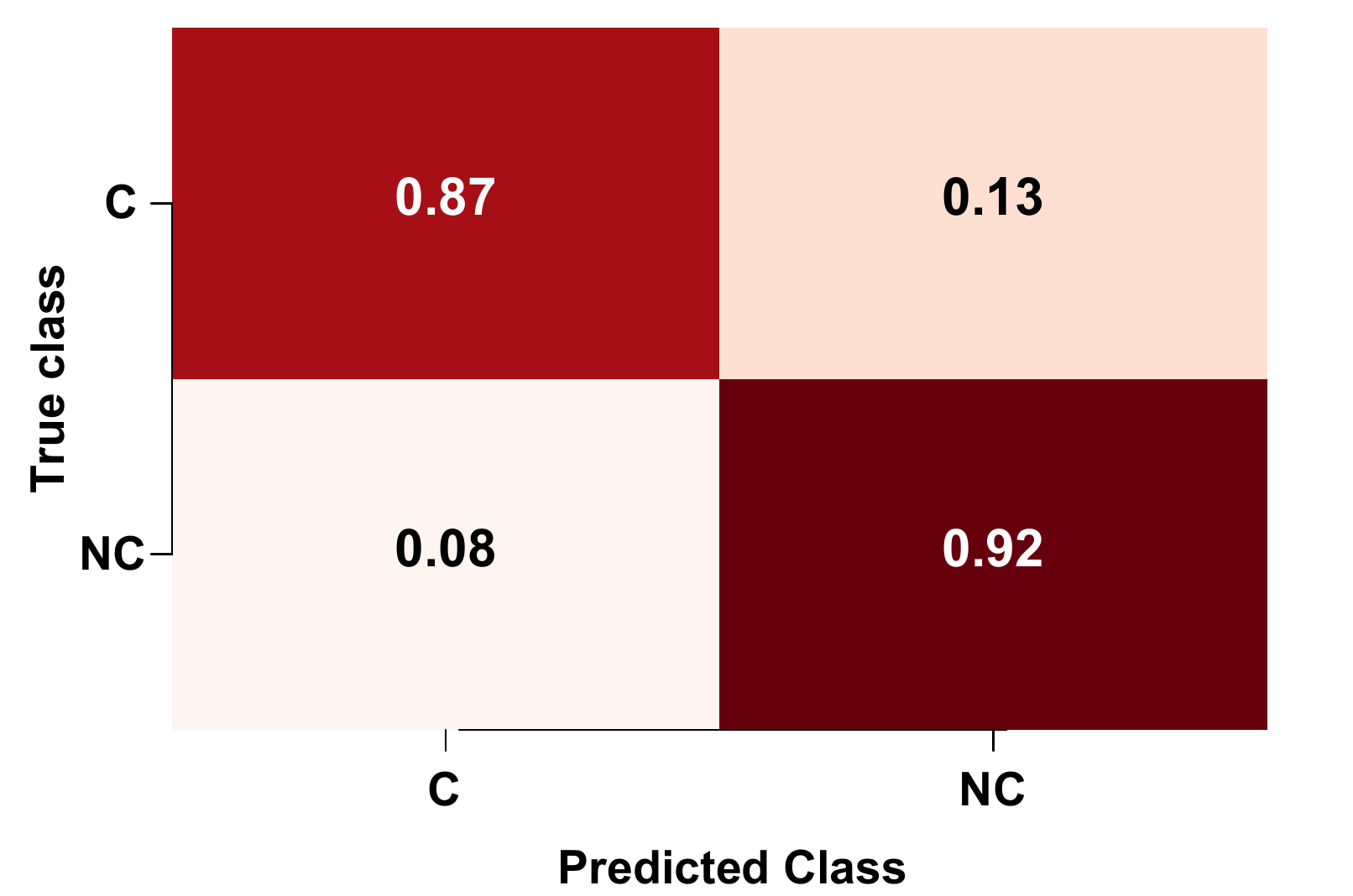}
    \end{center}
    \caption{Confusion Matrix for the complete classifier using the Test Set. Values are normalized with respect to the total number of contracts in each class.}
    \label{fig:ConfusionMatrix}
\end{figure}

Figure \ref{fig:ConfusionMatrix} shows the Confusion Matrix of the complete classification model using the Test Set as input data. The values of the matrix are normalized with respect to the number of contracts in each class to account for the imbalance between $C$ and $NC$ classes. The model correctly classified 87\% of the corrupt contracts, while 13\% were classified as non-corrupt. Conversely, the model correctly classified 92\% of the non-corrupt contracts and mistakenly classified 8\% as corrupt\footnote{To clarify, suppose that the classifier receives 1,000 new unknown contracts and that the $NC$:$C$ ratio remains at 45:1. In this example, the model will correctly classify 19 of the 22 corrupt and 900 of the 978 non-corrupt contracts. In total, 97 contracts would be predicted as corrupt, from which 19 would actually be. A posterior legal investigation would be needed to determine the legality of those 97 contracts. However, this result shows that with this model, the final set of contracts to focus the legal investigation was reduced from 1,000 to 97, missing only 3 corrupt contracts.}.

Evaluating the model's overall performance is difficult when the input data is as imbalanced as in our case. For example, a trivial model that classifies as $NC$ every input data would misclassify all the corrupt contracts and still achieve an overall accuracy of 0.98. For this reason, we calculated different metrics to evaluate the performance of our model, presented in the Table \ref{tab:metrics}. Particularly useful is the Balanced Accuracy (BAcc=0.89), defined as the average of the accuracy in each class, which takes into account the accuracy for $C$ and $NC$ independently.

\begin{table*}[htb!]
\begin{tabularx}{\textwidth}{l@{\hspace{1em}}l@{\hspace{1em}}l@{\hspace{1em}}l@{\hspace{1em}}X}
\hline \hline
\textbf{Metric}   & \textbf{Value} & \textbf{Range} & \textbf{Definition} & \textbf{Description} \\
\hline
Accuracy          & 0.92    &  0.0 - 1.0       &     Acc= CP/TP        & The proportion of correct predictions (CP) out of the total number of predictions (TP). \\
Balanced Accuracy &     0.89           &    0.0 - 1.0        & BAcc= (TPR + TNR)/2      & The mean value between the true positive rate (TPR) and the true negative rate (TNR). It accounts for unbalanced data.  \\
$NC$ Accuracy & 0.92 & 0.0 - 1.0 & NC Acc = NC$_{CP}$/$|$NC$|$ & The per-class accuracy of $NC$ class, defined as the proportion of correctly classified $NC$ contracts (NC$_{CP}$) out of the total number of $NC$ contracts $|$NC$|$. \\

$C$ Accuracy & 0.87 & 0.0 - 1.0 & C Acc = C$_{CP}$/$|$C$|$ & The per-class accuracy of $C$ class, defined as the proportion of correctly classified $C$ contracts (C$_{CP}$) out of the total number of $C$ contracts $|$C$|$.  \\
AUC               &       0.94         &    0.0 - 1.0    &  N/A & The Area Under the Curve aggregates the performance of a binary classifier (True Positive Rate (TPR) vs False Positive Rate(FPR)) on all possible threshold values. It is calculated as the total area under the associated ROC curve.  \\
Precision         & 0.99    &  0.0 - 1.0    &   P=TP/(TP + FP)    &    The proportion of true positive predictions (TP) out of the total of positive predictions (True Positive $+$ False Positive (TP $+$ FP)).   \\
Recall            & 0.92    &  0.0 - 1.0       &    R=TP/(TP + FN) &    The fraction of samples from a class which are correctly predicted by the model, calculated as the true positive predictions (TP) out of the total samples in a class (TP + FP). \\
F1-Score          & 0.95 &    0.0 - 1.0      & F1=2$\frac{{\mbox {P*R}}}{{\mbox {P+R}}}$   & The harmonic mean between precision (P) and Recall (R). It combines both metrics in a single measure but ignores the true negatives. \\
\hline \hline
\end{tabularx}
\caption{{\it Model results - All features.} Standard performance measures of the machine learning system for public procurement. Here, the positive cases refer to Non-Corrupt contracts. In all cases, the performance of the system is above 0.9. N/A, Not Applicable. All definitions were taken from \citep{alpaydin2020,geron2019,kuhn2013,mitchell1997}}
\label{tab:metrics}
\end{table*}

\begin{figure*}[htb!]
    \begin{center}
        \includegraphics[width=0.95\textwidth]{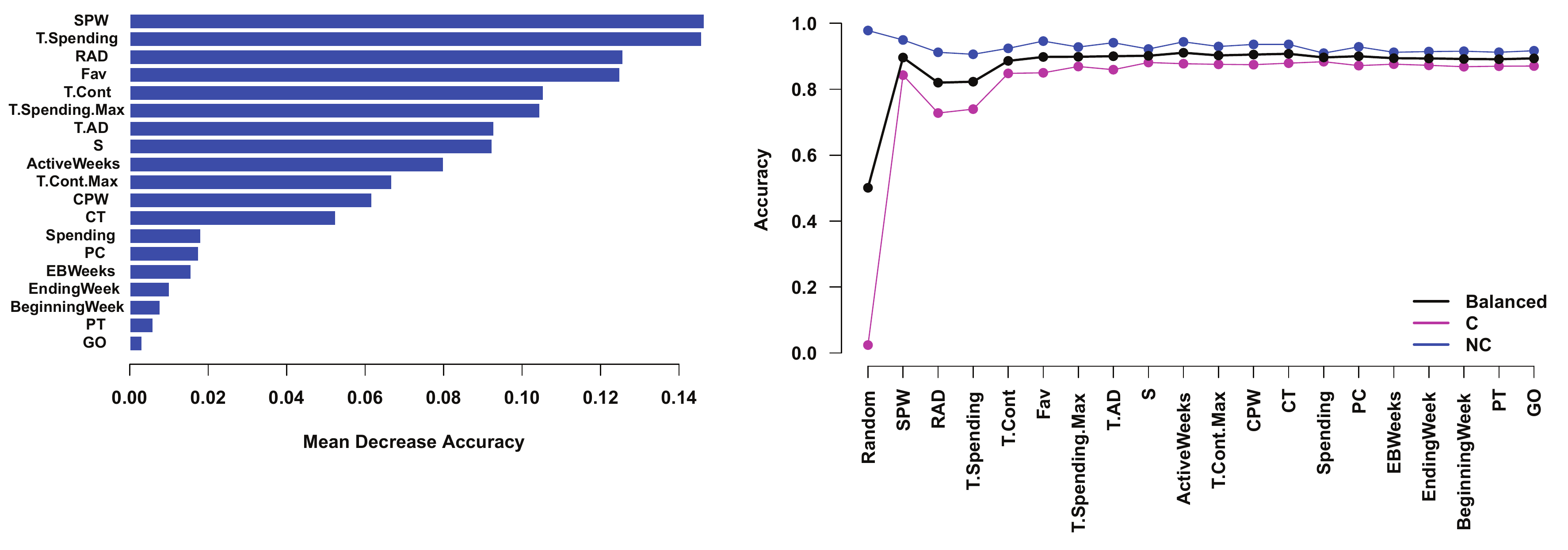}
    \end{center}
    \caption{{\it Left}: Feature importance analysis based on the random forest classifiers during training. The importance of each variable is measured as the mean decrease in accuracy under permutation of the values of that variable. {\it Right}: Recursive Feature Elimination analysis; for each feature on the $x$-axis, the corresponding points indicate the accuracy of the complete classification system before eliminating that feature. The balanced accuracy is the average of the accuracy for the classes $NC$ and $C$.
}
    \label{fig:RecFeatElim}
\end{figure*}

We next explored the relative importance of the features as predictors in our model. To determine the most relevant variables, we calculated the importance of the variable as described in section \ref{sec:importancevariable}. As  Figure \ref{fig:RecFeatElim}-Left shows, variables {\bf Spending} to {\bf GO} have little effect on the system's performance, as reflected in a mean decrease of accuracy lower than 0.02. Remarkably, the only variables of type i) (those solely describing characteristics of the contract) with a mean decrease accuracy larger than 0.02 are Contract Type ({\bf CT}) and Stratification ({\bf S}). Conversely, the most important variables are type ii)-iv), reflecting either a risk factor (type iv))  or the relationship between buyer and supplier (type ii)). The risk factor {\bf SPW} and the buyer-supplier relationship {\bf T.Spending} are the features with the highest mean decrease in accuracy. 

Finally, we applied Recursive Feature Elimination, as described in section \ref{sec:RFE}, to eliminate non-relevant variables for the classification. Figure \ref{fig:RecFeatElim}-Right shows the relative accuracy of $C$ and $NC$ classes and their balanced accuracy in each stage of RFE. For clarity, we present the results in the forward direction instead of the backward process performed by RFE. Each point's triplet corresponds to the classifier's accuracy using all the previous variables. The first triplet corresponds to a random classifier that predicts $C$ or $NC$ with a probability proportional to the number of corrupt or non-corrupt contracts in the entire dataset. Table \ref{tab:recFeatEl} shows the numerical results in Figure \ref{fig:RecFeatElim}-Right. The collection of features that maximizes the Balanced$=0.91$, NC$=0.94$, and C$=0.88$ accuracies includes the variables {\bf SPW} to {\bf CT}, which is consistent with the variable importance of Figure \ref{fig:RecFeatElim}-Left. 

\begin{table}[htb!]
\begin{tabular}{lcccc}
\hline \hline
\textbf{Feature} & \textbf{Type} & \textbf{Balanced} & \textbf{NC} & \textbf{C} \\
& &{\bf Accuracy} & {\bf Accuracy} &{\bf Accuracy} \\
\hline
Random & N/A & 0.50 &  0.98  & 0.02                \\
\bf SPW  & \bf iv & \bf 0.90   & \bf 0.95  & \bf 0.85   \\
\bf RAD  & \bf iv & \bf 0.82 & \bf 0.91 & \bf 0.72      \\
\bf T.Spending &  \bf ii & \bf 0.82 & \bf 0.89 & \bf 0.75 \\
\bf T.Cont & \bf ii  &\bf 0.89 & \bf 0.93 & \bf 0.84    \\
\bf Fav & \bf iv& \bf 0.90 & \bf 0.95 & \bf 0.86 \\
\bf T.Spending.Max & \bf iii & \bf 0.90 & \bf 0.94 &\bf 0.86 \\
\bf T.AD & \bf ii & \bf 0.90 & \bf 0.94 & \bf 0.87 \\
\bf S & \bf i & \bf 0.91 & \bf 0.94  & \bf 0.87 \\
\bf ActiveWeeks & \bf ii & \bf 0.90 & \bf 0.93  & \bf 0.88 \\
\bf T.Cont.Max & \bf iii & \bf 0.91 & \bf 0.94 & \bf 0.87  \\
\bf CPW & \bf iv & \bf 0.91 & \bf 0.93 & \bf 0.88 \\
\bf CT  & \bf i  & \bf 0.91 & \bf 0.94 & \bf 0.88 \\
Spending & i & 0.90 & 0.92 & 0.88 \\
PC & i & 0.90 & 0.93 & 0.87 \\
EBWeeks & i & 0.90 & 0.92 & 0.87    \\
EndingWeek & i & 0.90    & 0.91    & 0.88 \\
BeginningWeek & i & 0.89 & 0.91    & 0.87  \\
PT  & i & 0.89  & 0.91  & 0.87 \\
GO  & i & 0.89  & 0.91  & 0.87       \\
\hline \hline
\end{tabular}
\caption{Recursive Feature Elimination analysis corresponding to Fig.\ref{fig:RecFeatElim}-Right. Bold face represent the set of features with the best results for $C$ and $NC$ contracts, with a C- Accuracy = 0.88, NC-Accuracy = 0.94 and a Balanced-Accuracy=0.91}
\label{tab:recFeatEl}
\end{table}

\begin{figure}[htb!]
\begin{center}
\includegraphics[width=0.5\textwidth]{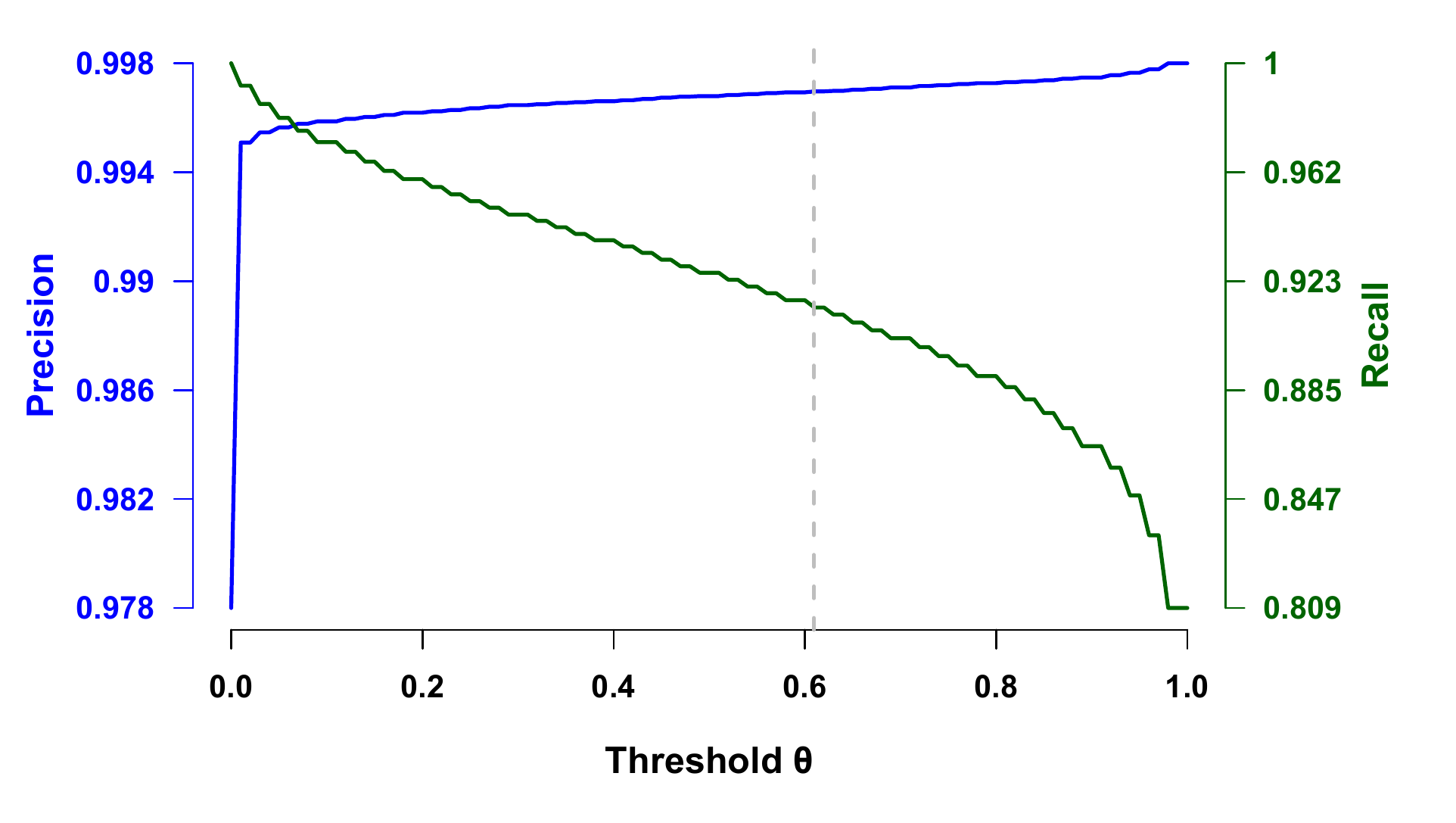}
\end{center}
\caption{Effect of changing the corruption threshold $\theta$ in precision and recall. The gray dashed vertical line indicates the threshold $\theta=0.61$ that optimized the trade-off between TPR and FPR.}
\label{fig:s_precrec}
\end{figure}

\begin{figure*}[htb!]
\begin{center}
\includegraphics[width=0.85\textwidth]{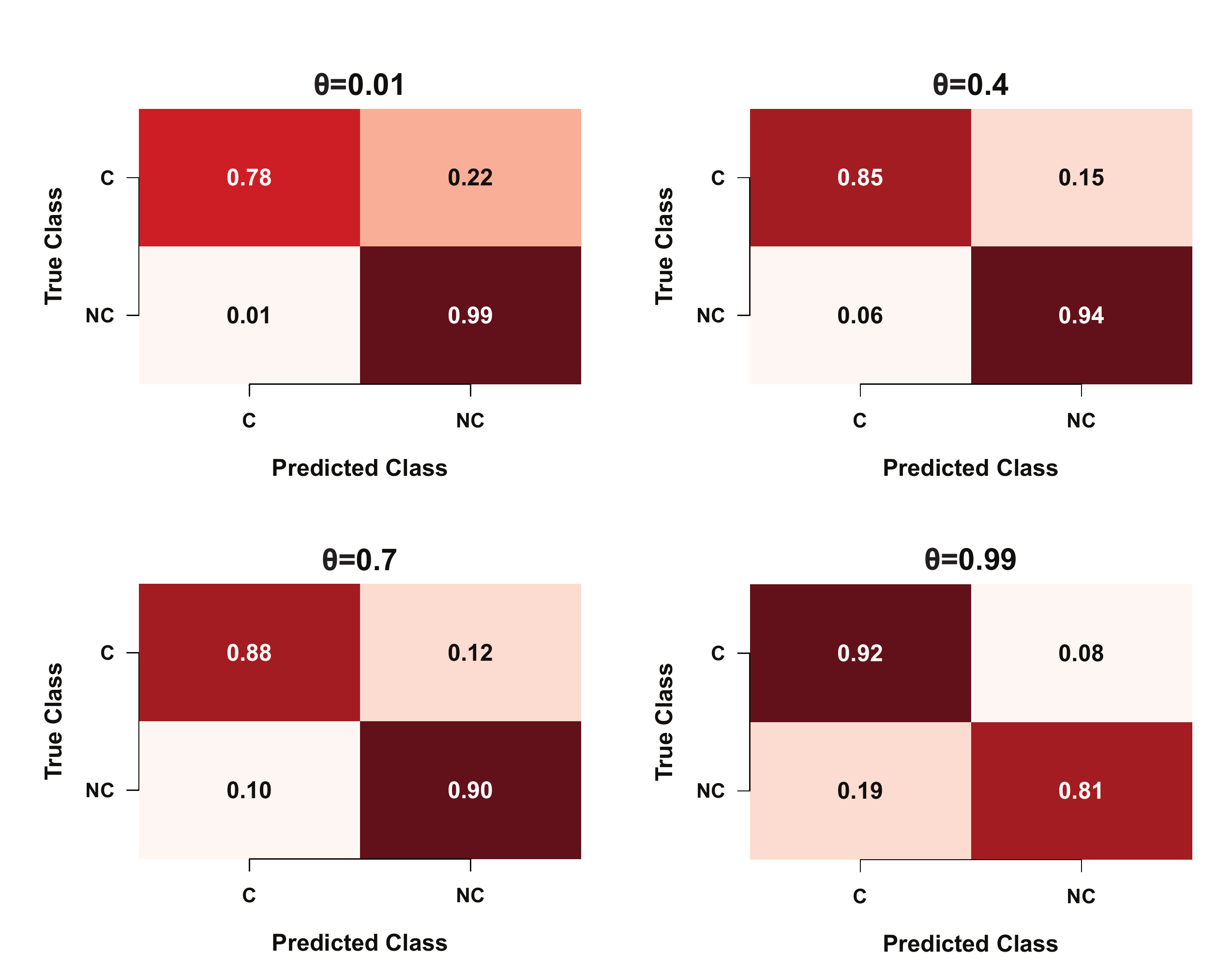}
\end{center}
\caption{Confusion Matrices with different values of the corruption threshold $\theta$. Lower values of $\theta$ increase the accuracy in detecting $NC$ contracts at the cost of reducing the accuracy of $C$ ones. Higher values of $\theta$ increase the accuracy in detecting $C$ contracts while increasing the number of $NC$ contracts misclassified as $C$.}
\label{fig:s_Matrices}
\end{figure*}

\section{Discussion}

The usual procedures for detecting corruption in public procurement may require physically visiting the supplier involved, auditing the company, matching receipts to the services or products provided, verifying that the specifications of the contract are actually met, and so on. These efforts usually consume a large amount of time and resources, thus, selecting which contracts or companies to investigate is crucial. This selection may be done in several ways: companies may be chosen at random; they may incur in suspicious behavior, such as delays; they may be reported by whistle blowers or journalists, etc. To complement these selection methods, attempts have been made to take advantage of the large amount of data that can be accessed nowadays, to use statistical methods and diverse machine learning tools to try to identify and characterize corruption in different stages of public procurement.  Here we propose a machine learning model to predict corruption in public procurement, trained with Mexican public contracts from 2013 to 2020, achieving a final balanced accuracy of 91\%. Our system is trained to detect potential corrupt contracts, using variables from different types as predictors, representing the individual contract properties, the buyer-supplier relationship, the buyer's features and corruption risk factors. 

The training process of our model includes labeling each contract as Corrupt ($C$) or Non-Corrupt ($NC$) according to whether or not the supplier had been found guilty of corruption in regards to any public contract he had participated in, either since or before the assignment of the contract being labeled. The characteristics of the data may impose a significant challenge on any machine learning model; for our study case, we should remark that corrupt contracts acquire that label only after an investigation from the Mexican Federal Government, but some of the contractors are appealing this decision, and some others are concluding their investigations, so that the list of corrupt contracts may change over time. On the other hand, the $NC$ label was assigned following the principle of presumption of innocence; thus, we are aware that some corrupt contracts went undetected and were misclassified as non-corrupt; however, we expect that their statistical weight was minimum. Therefore, and because of the label changes as a consequence of future legal investigations, our model would benefit as new improved data about corrupt and non-corrupt contractors becomes available. 

%we are sure only on the label of corrupt contracts??. We should remark that corrupt contracts acquire that label only due to an investigation from the Mexican Federal Government. However, non-corrupt contracts (the vast majority) have not been under such an investigation O SÍ, meaning that we are not sure if a non-corrupt contract is an honest one or if it just has not been under review yet. Our model would greatly benefit from new data on proven non-corrupt contracts.} \\

Each machine learning algorithm requires a particular pre-processing of the data, and various algorithms may show different performances. The machine learning model presented here is grounded on the Random Forest. We choose this algorithm given its capacity to deal naturally with categorical and numerical variables, detect the most relevant features used for the classification and work with an ensemble of classifiers, reducing the probability of overfitting. The present model is flexible enough to adapt to machine learning classification algorithms, other than Random Forest, with minor changes in the data pre-processing. This flexibility and modularity allows comparing different classifiers using feature subsets or changing the dataset entirely.

The dataset is highly imbalanced, with a $C:NC$ ratio of $1:45$. Imbalanced data may negatively affect the performance of many machine learning algorithms, including Random Forest. To contend with this effect, we use Repeated Random Sub-sampling to implement a voting system where each voter is a Random Forest, trained with a balanced sub-sample, we called this model an Hyper-Forest. The final result is determined by whether or not the number of voters is above a threshold, determined during the calibration stage by a ROC curve. This process achieves a balanced accuracy of 89\%, which was improved to 91\% using Recursive Feature Elimination, with relative accuracies of 88\% and 94\% for $C$ and $NC$ classes, respectively. Overall, the combined application of Repeated Random Sub-sampling, ROC calibration and  Recursive Feature Elimination was the best approach to classify the imbalanced data presented in our study.

The Importance of the Variables and Recursive Feature Elimination prove that the most relevant variables for classification in the model are related to either the buyer/supplier relationship (variables of type ii)) or risk factors (variables of type iv)). At the same time, all the variables excluded from the best set of features are individual contract properties (variables of type i)). In fact, removing such predictors improved the classifier's performance. Our results suggest that corruption is not a practice in individual contracts; instead, it is a systematic behavior between buyers and suppliers. 

%For this reason, in order to find corrupt practices in public procurement, efforts should be made to investigate the relationship between buyers and suppliers rather than individual contracts. 

In spite of the efficiency obtained by our model, we should be aware of the limits of these kinds of methods because even when they can represent a considerable advance in the fight against corruption, they are not a definitive solution. First of all, the patterns identified as corrupt are based on those already recognized by human experience, which means that the system is not trained to identify new patterns. This may move criminals to devise alternative corruption strategies that would go undetected. Thus, the model needs to be fed with new data to stay up to date and useful. Second, the model may misidentify straight contracts as corrupt. Indeed, even when the system can be calibrated to make some compromises between different errors, there is always a failure rate to consider. Then, these predictor methods can be used to identify and prioritize potential corrupt contracts among thousands, but human intervention will always be necessary to make the final decision. 

Our main goal was to implement a machine learning model to recognize corrupt contracts, maximizing the trade-off between correctly classifying non-corrupt contracts (TPR) and misidentifying as few corrupt contracts as possible (FPR). However, the model can be re-calibrated for other criteria by selecting different values of the corruption threshold $\theta$ (section \ref{sec:model}). Figures \ref{fig:s_precrec} and \ref{fig:s_Matrices}  show the effects of taking different values of $\theta$ in Precision, Recall and the Confusion Matrix. Reducing $\theta$ to a value closer to 0 would reduce the number of $NC$ contracts erroneously identified as corrupt and increase the $NC$ contracts correctly classified (increasing Recall), but also increase the corrupt contracts misclassified as $NC$ (decreasing Precision). On the other hand, increasing $\theta$ to a value closer to 1 would reduce the number of corrupt contracts misclassified as $NC$ (increasing Precision) and increase the $C$ contracts correctly classified, but also increase the $NC$ contracts misclassified as corrupt (reducing Recall).

The methods and tools described here are general enough to be implemented to train a machine learning model with the data available from different countries, even with different features, with the only restriction that the contracts must be labeled as $C$ and $NC$. Finally, our work presents a tool that can be easily implemented by agencies and governments to help in the decision-making process to identify, predict and analyze corruption in public procurement contracts.

%Besides, our results show that machine learning tools implemented here can be used to identify corrupt contracts in public procurement among big data sets.

%Our results show that machine learning tools can be used to identify corrupt contracts among a big data set with similar accuracy to that performed by humans \textcolor{red}{A QUE TE REFIERES???} but in much less time and effort. 
%Our goal was to train a Random Forest model to recognize corrupt contracts, while misidentifying as few non-corrupt contracts as possible. However, the algorithm can be re-calibrated for other criteria, for example, to reduce the number of falsely identified corrupt contracts, at the cost of missing more corrupt ones, or viceversa. The model was trained with the presently available data regarding the public procurement contracts made by the Mexican governments from 2013 to 2020, however, we show that the application of recursive variable elimination may be used to test the importance of newly added variables, and analyze whether they are significant or not for the system's performance.\\

% We have illustrated the potential of machine learning models to help detect corruption in public procurement by exploiting the access to big administrative data sets. 

\section{Acknowledgements}
AFC thanks PostDoctoral Scholarship DGAPA-UNAM for financial support.

\vspace{1.0cm}
\FloatBarrier

\def\bibsection{}  
\centerline{\textbf{\normalsize 7. REFERENCES}}
\bigbreak

%\bibliography{bilbio.bib}
\bibliographystyle{dcu}

\end{document}